\documentclass[showkeys,aps,10pt,twocolumn,showpacs,preprintnumbers,amsmath,amssymb,prd,letterpaper,floatfix,nofootinbib,superscriptaddress,]{revtex4-1}
\usepackage{graphics}
\usepackage{graphicx}
\usepackage{epsfig}
\usepackage[usenames]{xcolor}
\usepackage{longtable}
\PassOptionsToPackage{hyphens}{url}\usepackage{hyperref}
\usepackage{soul}
\usepackage{breakurl}
\usepackage{latexsym}
\usepackage{amsmath}
\usepackage{amsthm}
\usepackage{amsfonts}
\usepackage{amssymb}
\usepackage{dsfont}
\usepackage{bm} 
\newcommand{\be}{\begin{equation}}
\newcommand{\ee}{\end{equation}}
\newcommand{\bel}{\begin{align}}
\newcommand{\eel}{\end{align}}
\newcommand{\bea}{\begin{eqnarray}}
\newcommand{\eea}{\end{eqnarray}}

\begin{document}

\title{ \texorpdfstring{Pion-nucleon scattering in the Roper channel  from lattice QCD  }{}}

\author{C.~B.~Lang}
\email{christian.lang@uni-graz.at}
\affiliation{Institute of Physics,  University of Graz, A--8010 Graz, Austria}

\author{L.~Leskovec}
\email{leskovec@email.arizona.edu}
\affiliation{Department of Physics, University of Arizona, Tucson, AZ 85721, USA} 

\author{M.~Padmanath}
\email{Padmanath.Madanagopalan@physik.uni-regensburg.de}
\affiliation{Institute of Physics,  University of Graz, A--8010 Graz, Austria}
\affiliation{Institut f\"ur Theoretische Physik, Universit\"at Regensburg, D-93040 Regensburg, Germany}

\author{S. Prelovsek}
\email{sasa.prelovsek@ijs.si}
\affiliation{Department of Physics, University of Ljubljana, 1000 Ljubljana, Slovenia}
\affiliation{Jozef Stefan Institute, 1000 Ljubljana, Slovenia}
\affiliation{Institut f\"ur Theoretische Physik, Universit\"at Regensburg, D-93040 Regensburg, Germany}
\affiliation{Theory Center, Jefferson Lab, 12000 Jefferson Avenue, Newport News, Virginia 23606, USA}

 \date{\today}

\begin{abstract} 
We present a lattice QCD study of $N\pi$ scattering in  the positive-parity nucleon
channel, where the puzzling Roper resonance $N^*(1440)$ resides in experiment.  The study 
is based on the PACS-CS ensemble of gauge configurations with $N_f=2+1$ Wilson-clover
dynamical fermions, $m_\pi \simeq 156~$MeV and $L\simeq 2.9~$fm.  In addition to a number
of $qqq$ interpolating fields, we implement operators for $N\pi$ in $p$-wave and $N\sigma$
in $s$-wave. In the center-of-momentum frame we find three eigenstates below 1.65 GeV.
They are dominated by $N(0)$, $N(0)\pi(0)\pi(0)$ (mixed with $N(0)\sigma(0)$) and  $N(p)\pi(-p)$ with $p\simeq 2\pi/L$,
where momenta are given in parentheses. This is the first  simulation where the expected
multi-hadron states are found in this channel. The experimental $N\pi$ phase-shift would
--  in the approximation of purely elastic $N\pi$ scattering -- imply an additional 
eigenstate near the Roper mass $m_R\simeq 1.43~$GeV for our lattice size. We do not
observe 	 any such additional  eigenstate, which indicates that $N\pi$ elastic
scattering alone does not render a low-lying Roper. Coupling with other channels, most
notably with $N\pi\pi$, seems to be important for generating the Roper resonance,
reinforcing the notion that this state could  be a  dynamically generated resonance.  Our
results are in line with most of previous lattice studies based just on $qqq$
interpolators, that did not find a Roper eigenstate below $1.65~$GeV.   The study of the 
coupled-channel scattering  including a three-particle decay $N\pi\pi$ remains a
challenge.  
 \end{abstract}
 

\maketitle

\section{Introduction}

Pion-nucleon  scattering  in the $J^{P}=1/2^{+}$ channel captures the information on the
excitations of the nucleon ($N=p, n$).  The $N\pi$ scattering in $p$-wave  is elastic only
below the inelastic threshold  $m_N+2m_\pi$ for $N\pi\pi$. The main  feature   in this
channel at low energies is the so-called Roper resonance with $m_R=(1.41-1.45)~$GeV and
$\Gamma_R=(0.25-0.45)~$GeV  \cite{pdg14} that was first introduced by L.D. Roper 
\cite{Roper:1964zza} to describe the experimental $N\pi$ scattering. The resonance decays
to $N\pi$ in $p$-wave with a branching ratio $Br\simeq 55-75\%$ and to $N\pi\pi$ with
$Br\simeq 30-40\%$ (including  $N(\pi\pi)^{I=0}_{s-wave}$, $\Delta\pi$ and $N\rho$), while
isospin-breaking and electromagnetic decays lead to a $Br$ well below one percent.

Phenomenological  approaches that considered the $N^*(1440)$ resonance as dominantly $qqq$
state, for example quark models \cite{Isgur:1978xj,Liu:1983us,Capstick:1986bm}, gave a
mass that is too high and a width that is too small in comparison to experiment. This led
to several suggestions on its nature and a large number of phenomenological studies. One
possibility is a dynamically generated Roper resonance where the coupled-channel
scattering $N\pi/N\sigma/ \Delta\pi$  describes the $N\pi$ experimental scattering data 
without any excited $qqq$ core
\cite{Krehl:1999km,Schutz:1998jx,Liu:2016uzk,Matsuyama:2006rp}.  The scenarios with
significant $qqqq\bar q$ Fock components  \cite{Jaffe:2004zg,JuliaDiaz:2006av} and hybrids
 $qqqG$  with gluon-excitations \cite{Golowich:1982kx,Kisslinger:1995yw} were also
explored. The  excited $qqq$ core, where the interaction of quarks is supplemented by the
pion exchange,   brings the mass closer to experiment
\cite{Glozman:1995fu,Glozman:1997ag}.  A similar effect is found as a result of some other
mechanisms that accompany the $qqq$ core, for example a vibrating $\pi \sigma$
contribution \cite{Alberto:2001fy} or coupling to all allowed channels
\cite{Golli:2007sa}. These  models are not directly based on QCD, while  the effective
field theories contain a large number of low-energy-constants  that need to be determined
by other means. The rigorous Roy-Steiner approach is based on phase shift data and
dispersion relations implementing unitarity, analyticity and crossing symmetry; it leads
to $N\pi$ scattering amplitudes at energies $E\leq 1.38~$GeV that do not cover the whole
region of the Roper resonance \cite{ Hoferichter:2015hva}.  The implications of the
present simulation on various  scenarios  are discussed in Section \ref{sec:discussion}.

All previous lattice QCD simulations, except for \cite{Kiratidis:2016hda},  addressed excited states in
this channel  using three-quark operators; this has conceptual issues for a strongly decaying resonance 
where coupling to multi-hadron states is essential. In principle multi-hadron eigenstates can also arise 
from the  $qqq$ interpolators in a dynamical lattice QCD simulation but in practical calculations the 
coupling to $qqq$ was too weak for an effect. Another assumption of the simple operator approach is 
that the energy of the first excited eigenstate is identified with the mass of  $N^*(1440)$, which is a drastic 
approximation for a wide resonance. The more rigorous L\"uscher approach 
\cite{Luscher:1990ux,Luscher:1991cf} assuming elastic scattering predicts an eigenstate  in the energy region within the resonance width  (see Fig.  \ref{fig:E_analytic}). 

The masses of  the Roper obtained in the recent dynamical lattice simulations
\cite{Liu:2014jua,Alexandrou:2013fsu,Alexandrou:2014mka,Engel:2013ig,Edwards:2011jj,Mahbub:2013ala,Roberts:2013ipa} using the $qqq$ approach  are summarized in 
\cite{Leinweber:2015kyz}. Extrapolating these to physical quark masses, where
$m_{u/d}\simeq m_{u/d}^{phys}$, the Roper mass was found above $1.65~$ GeV by all
dynamical studies except \cite{Liu:2014jua}, so most of the studies disfavour a low-lying
Roper $qqq$ core. The only dynamical study that observes a mass around $1.4~$GeV was done
by  the $\chi$QCD  collaboration \cite{Liu:2014jua}; it was based on the fermions with
good chiral properties (domain-wall sea quarks and overlap valence quarks) and employed a 
Sequential Empirical Bayesian (SEB) method to extract eigenenergies from a single
correlator. It is not yet finally settled
\cite{Leinweber:2015kyz,Liu:2014jua,Liu:2016rwa,chuan_lat16} whether the discrepancy of
\cite{Liu:2014jua} with other results is related to the chiral properties of quarks, use
of SEB   or poor variety of interpolator spatial-widths in some studies\footnote{The
$\chi$QCD collaboration \cite{Liu:2016rwa}  recently verified   that SEB and variational
approach with wide smeared sources ($r\simeq 0.8~$fm)  lead to compatible   $E\simeq
1.9~$GeV for Wilson-clover fermions and $m_\pi\simeq 400~$MeV.}. Linear combinations of
operators with different spatial widths allow to form the radially-excited eigenstate with
a node in the radial wave function, which was found at $r\simeq 0.8~$fm in
\cite{Roberts:2013ipa,Roberts:2013oea,Liu:2014jua}.

An earlier quenched  simulation \cite{Mathur:2003zf} based on $qqq$ interpolators 
used overlap fermions and  the SEB method to extract eigenenergies.  The authors find a
crossover between first excited $1/2^+$ state and ground $1/2^-$ state as a function of
the quark mass, approaching the experimental situation.  A more recent quenched
calculation \cite{Mahbub:2009aa} using FLIC fermions with improved chiral properties  and
variational approach also reported a similar observation.

In continuum the $N^*(1440)$ is not an asymptotic state
but a strongly decaying resonance that manifests itself in the continuum of $N\pi$ and
$N\pi\pi$ states. The spectrum of those states becomes discrete on the finite
lattice of size $L$. For non-interacting $N$ and $\pi$ the periodic boundary conditions in
space  constrain the momenta  to   multiples of $2\pi/L$. The interactions modify the
energies of these discrete multi-hadron states and possibly render additional eigenstates.

The multi-hadron states have never been established  in the previous lattice simulations
of the Roper channel, although they should inevitably appear as eigenstates in dynamical
lattice QCD. In addition to being important representatives of the $N\pi$ and $N\pi\pi$
continuum, their energies and number in principle provide  phase shifts for the scattering
of nucleons and pions. These, in turn,  provide information on the Roper resonance that
resides in this channel. In the approximation when $N\pi$ is decoupled from other channels
the $N\pi$ phase shift and the scattering matrix are directly related to eigenenergies via
the L\"uscher method \cite{Luscher:1990ux,Luscher:1991cf}.  The determination of the
scattering matrix for coupled two-hadron channels has been proposed in
\cite{Doring:2011vk,Hansen:2012tf} and was recently extracted from a lattice QCD
simulation \cite{Dudek:2014qha,Dudek:2016cru} for other cases.  The presence of the
three-particle decay mode $N\pi\pi$  in the Roper channel, however,  poses a significant
challenge to the rigorous treatment, as the scattering matrix for three-hadron decay has
not been extracted from the lattice yet, although impressive progress on the analytic side
has been made \cite{Hansen:2015zga}.

The purpose of the present paper is to determine the  complete discrete spectrum for the
interacting system with $J^P=1/2^+$, including  multi-hadron eigenstates. Zero total
momentum is considered since parity is a good quantum number in this case. In addition to
$qqq$ interpolating fields, we incorporate for the first time   $N\pi$ in $p$-wave in
order to address their scattering. The  $N\sigma$ in $s$-wave is also employed to account
for $N(\pi\pi)^{I=0}_{s-wave}$.  We aim at the energy region below $1.65~$GeV, where the
Roper resonance is observed in experiment. In absence of meson-meson and meson-baryon  interactions one expects eigenstates dominated by $N(0)$,  $N(0)\pi(0)\pi(0)$, $N(0)\sigma(0)$ and $N(1)\pi(-1)$,  in our $N_f=2+1$
dynamical simulation for $m_\pi\simeq 156~$MeV and $L\simeq 2.9~$fm.   The momenta in
units of $2\pi/L$ are given in parenthesis. $N$ and $\pi$ in $N\pi$ need at  least
momentum $2\pi/L$  to form the $p$-wave. The  PACS-CS configurations \cite{Aoki:2008sm}
have favourable parameters  since the  non-interacting energy
$\sqrt{m_\pi^2+(2\pi/L)^2}+\sqrt{m_N^2+(2\pi/L)^2}\simeq 1.5~$GeV of $N(1)\pi(-1)$ falls
in the Roper region. The number of observed eigenstates and their energies   will lead to
certain implications concerning the Roper resonance.

In the approximation of elastic $N\pi$ scattering, decoupled from $N\pi\pi$, the
experimentally measured $N\pi$ phase shift predicts  four eigenstates below $1.65~$GeV, as
argued in Section \ref{sec:discussion_elastic} and  Figure \ref{fig:E_analytic}. Further
analytic guidance for this channel was recently presented in  \cite{Liu:2016uzk}, where
the  expected discrete lattice spectrum (for our $L$ and $m_\pi$)  was  calculated using a
Hamiltonian Effective Field Theory (HEFT) approach for three hypotheses concerning the
Roper state (Fig. \ref{fig:HEFT}). All scenarios involve channels 
$N\pi/N\sigma/\Delta\pi$ (assuming stable $\sigma$ and $\Delta$) and are apt to reproduce
the experimental $N\pi$ phase shifts. The scenario which involves also a bare Roper $qqq$
core predicts four eigenstates in the region $E<1.7~$GeV of our interest, while the
scenario without Roper $qqq$ core predicts three eigenstates
\cite{Liu:2016uzk}.\footnote{This numbering omits the $\Delta(1) \pi(-1)$ and $N(1)\sigma(-1)$
eigenstates that are near $1.7~$GeV; these  are not  expected to be found in our study 
since the corresponding interpolators are not included. Our notation implies projection of all operators to $J^P=\frac{1}{2}^+$.} The Roper resonance in the second case is
dynamically generated  purely from the  $N\pi/N\sigma/\Delta\pi$ channels, possibly
accompanied by the ground state nucleon $qqq$ core.

As already mentioned, our aim is to establish the   expected low-lying multi-particle
states in the positive-parity nucleon channel. This has been already accomplished in the
negative-parity channel, where $N\pi$ scattering in $s$-wave was simulated in
\cite{Lang:2012db}. An exploratory study  \cite{Verduci:2014csa} was done in a moving
frame, where both parities contribute to the same irreducible representation. The only
lattice simulation in the positive-parity channel  that included (local) $qqqq\bar q$
interpolators in addition to $qqq$ was recently presented in  \cite{Kiratidis:2016hda}. 
No energy levels were found
between $m_N$ and $\simeq 2~$GeV for $m_\pi\simeq 411~$MeV.    The levels related to
$N(1)\pi(-1)$ and $N(0)\sigma(0)$ were not observed, although they are  expected below
$2~$GeV according to \cite{Liu:2016uzk}. This is possibly due to the local nature of the
employed $qqqq\bar q$ interpolators \cite{Kiratidis:2016hda}, which 
seem to couple too weakly to multi-hadron states in practice.

 This paper is organized as follows. Section \ref{sec:simulation} presents the ensemble,
 methodology, interpolators and other technical details to determine the eigenenergies.
 The resulting eigenenergies and  overlaps  are presented in Section \ref{sec:results},
 together with a discussion on the extraction of the $N\pi$ phase shift. The physics
 implications   are drawn in  Section \ref{sec:discussion} and an outlook is given in the
 conclusions.

\section{Lattice setup}\label{sec:simulation}
     
\subsection{Gauge configurations}\label{sec:conf}
     
We perform a dynamical calculation on 197  gauge configurations generated by the
PACS-CS collaboration with $N_f=2+1$, lattice spacing $a=0.0907(13)~$fm, lattice
extension $V=32^3\times 64$, physical volume $L^3\simeq (2.9~$fm$)^3$ and
$\kappa_{u/d}=0.13781$ \cite{Aoki:2008sm}. The quark masses, $m_{u}=m_d$, are nearly
physical and correspond to $m_\pi=156(7)(2)~$MeV as estimated by PACS-CS
\cite{Aoki:2008sm}. Our own estimate leads to somewhat larger $m_\pi$ as detailed
below (we still refer to it as an ensemble with  $m_\pi\simeq 156~$MeV). The quarks
are non-perturbatively improved Wilson-clover fermions, which do not respect exact
chiral symmetry (i.e., the Ginsparg-Wilson relation \cite{Ginsparg:1981bj}) at
non-zero lattice spacing $a$.  Most of the previous  simulations of the Roper channel
also employed Wilson-clover fermions, for example
\cite{Alexandrou:2013fsu,Alexandrou:2014mka,Edwards:2011jj,Mahbub:2013ala,Roberts:2013ipa}. 

Closer inspection of this ensemble reveals that there are a few configurations responsible
for a strong fluctuation of the pion mass,  which is listed in Table
\ref{tab:singlehadronmasses}. Removing one or four of the "bad" configurations changes the
pion mass by more than two  standard deviations. The configuration-set "all" indicates the
 full set of 197 gauge configurations, while "all-1" ("all-4") indicate a subset with 196
(193) configurations where one (four)  configuration(s) leading to the strong fluctuations
in  $m_\pi$ are removed\footnote{In the set 
\texttt{RC32x64\_B1900Kud01378100Ks01364000C1715} configuration  \texttt{jM000260} is
removed in "all-1", while  \texttt{jM000260, hM001460, jM000840} and \texttt{jM000860} are
removed in "all-4".}. 
\begin{table}[t]
\begin{ruledtabular}
\begin{tabular}{l  c  c }
config. set & $m_\pi $ [MeV] &$m_N$ [MeV]   \\
\hline
all   & $153.9 \pm 4.1 $  & $951\pm 19$ \\
all-1 & $163.9 \pm 2.4$  & $965 \pm13$ \\
all-4 & $164.4 \pm 2.1$  & $969 \pm 12 $  \\
\end{tabular}
\end{ruledtabular}
\caption{The single hadron masses obtained for the full ("all") set of configurations and
for the sets with one ("all-1") or four ("all-4") configurations omitted. Interpolators,
fit type and fit range are like in Table \ref{tab:singleH}. As discussed in the text our
final results are based on set "all-4". }\label{tab:singlehadronmasses}
\end{table}

We tested these three configuration-sets for a variety of hadron energies, and we find
that only $m_\pi$ varies  outside the statistical error, while  variations of masses for 
other hadrons  (mesons with light and/or heavy quarks and nucleon)   are smaller than the
statistical errors. This also applies for the nucleon mass  listed in     Table
\ref{tab:singlehadronmasses}. The energies of the pions and other hadrons with non-zero
momentum also do not vary significantly with this choice.

The Roper resonance is known to be challenging as far as statistical errors are concerned,
{especially for nearly physical quark masses}. The error on the masses and energies is
somewhat bigger for the full set than on the reduced sets in some cases, for example
$m_\pi$ and $m_N$ in Table \ref{tab:singlehadronmasses}. Throughout this paper, we will
present results for the reduced configuration-set "all-4", unless specified differently.
The final spectrum was studied for all three configuration-sets, and we arrive at the same
conclusions for all of them.

\subsection{Determining  eigenenergies}

We aim to determine the eigenenergies  in the Roper channel, and we will need  also the
energies of a single  $\pi$ or $N$.   Lattice computation of eigenenergies $E_n$  
proceeds by calculating the correlation matrix $C(t)$ for a set of interpolating fields
$O_{i}$($\bar O_{i}$) that annihilate (create) the physics system of interest
\begin{align}
C_{ij}(t)&=\langle \Omega|O_i(t+t_{src})\bar O_j (t_{src})|\Omega\rangle \nonumber \\
&= \sum_n \langle \Omega | O_i|n\rangle \mathrm{e}^{-E_nt} \langle n|\bar O_j |\Omega\rangle \nonumber \\
&= \sum_n  Z_i^n Z_j^{n*} \mathrm{e}^{-E_nt}
\label{C}
\end{align} 
with overlaps $Z_i^n=\langle \Omega | O_i|n\rangle$.   All our results are averaged over
all the source time slices $t_{src}=1,..,64$.

The $E_n$ and  $Z_j^{n}$ are extracted  from $C(t)$ via the generalized
eigenvalue method (GEVP) \cite{Michael:1985ne,Luscher:1985dn,Luscher:1990ck,Blossier:2009kd}
\begin{align} 
\label{gevp}
 C(t)u^{(n)}(t)&=\lambda^{(n)}(t)C(t_0)u^{(n)}(t)\;,\ \  \lambda^{(n)}(t)\propto  \mathrm{e}^{-E_n t}
\end{align} 
and we apply $t_0=2$ for all cases except for the single pion correlation where we choose $t_0=3$.
The large-time behavior of the eigenvalue  $\lambda^{(n)}(t)$ provides $E_n$, 
where specific fit forms will be mentioned case by case.   The  
\be
Z_j^{n}(t)=\mathrm{e}^{E_n t/2}  C_{jk}(t) u_k^{(n)}(t)/|C(t)^\frac{1}{2} u^{(n)}(t)| 
\label{eq:Z}
\ee 
give the overlap factors in the plateau region. 

For fitting $E_n$ from $\lambda^{(n)}(t)$ we usually employ a sum of two exponentials,
where the second exponential helps to parameterize the residual contamination from higher
energy states at small $t$ values.  For the single pion ground state we have a large range
of $t$-values to fit and there we combine $\cosh[E_n(t-N_T/2)]$ also with such an
exponential. Correlated fits are used throughout.   Single-elimination jackknife is used
for statistical analysis.

 \subsection{Quark smearing width and distillation}\label{sec:Dis}

\begin{figure}[!htb]
\begin{center}
\includegraphics*[width=0.45\textwidth,clip]{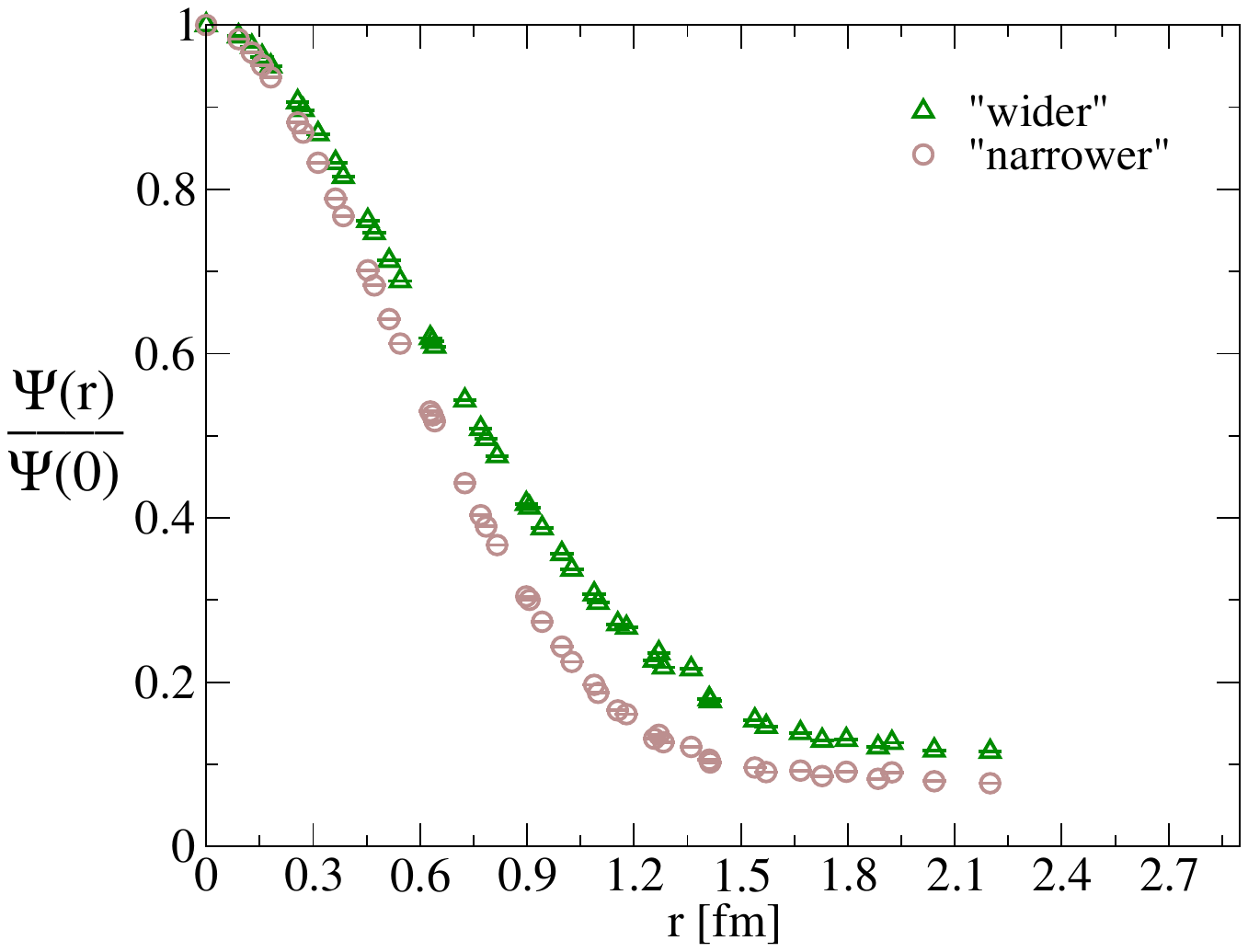}
\end{center}
\caption{ The profile $ \Psi(r)$ of the ``narrower" ($N_v=48$) and the ``wider" ($N_v=24$) 
smeared quark, where  $ \Psi(r)=\sum_{\mathbf{x},t}\sqrt{\mathrm{Tr}_c[~\square_{\mathbf{x,x+r}}(t)~
\square_{\mathbf{x,x+r}}(t)~]} $.}
\label{fig:smearing}
\end{figure}

The interpolating fields are built from the quark fields and we employ these with two
smearing widths illustrated in Fig. \ref{fig:smearing}.  Linear combinations of operators
with different smearing widths  provide  more freedom to form the eigenstates with nodes
in the radial wave function. This is favourable for the Roper resonance 
\cite{Roberts:2013ipa,Roberts:2013oea,Liu:2014jua}, which is a radial excitation within a
quark model.

Quark smearing is implemented using the so-called distillation method
\cite{Peardon:2009gh}. The method is versatile and enables us to compute all necessary
Wick-contractions, including terms with quark-annihilation. This is made possible  by
pre-calculating the quark  propagation  from specific quark  sources.  The sources are 
the lowest $k=1,..,N_v$ eigenvectors $v^{k}_{\mathbf{x}c}$ of the spatial lattice
Laplacian and $c$ is the color index. Smeared quarks are provided by  
$q^c(\mathbf{x})\equiv \square_{\mathbf{x'}c',\mathbf{x}c} \; q_{point}^{c'}(\mathbf{x'})$
\cite{Peardon:2009gh}  with the smearing operator
$\square_{\mathbf{x'}c',\mathbf{x}c}=\sum_{k=1}^{N_v}v^{k}_{\mathbf{x'}c'}v^{k\dagger}_{\mathbf{x}c}$. Different $N_v$ lead to different effective smearing widths.

In previous work we used stochastic  distillation \cite{Morningstar:2011ka} on this
ensemble, which is less costly but  renders noisier results. For the present project  we
implemented  the distillation\footnote{Sometimes referred to as the full distillation. }
with  narrower ($n$)  smearing $N_v=48$ and wider  ($w$) smearing   $N_v=24$, illustrated
in Fig. \ref{fig:smearing}.  Two smearings are employed to enhance freedom in forming the
eigenstates with nodes.  Most of the interpolators and results below are based on narrower
smearing which gives better signals in practice, although both widths are   not very
different.   The details of our implementation of the distillation  method are collected
in \cite{Lang:2011mn} for another ensemble.

\subsection{\texorpdfstring{Interpolators and energies of $\pi$ and $N$}{}}

Single particle energies are needed to determine reference energies of the non-interacting
(i.e., disregarding interaction between the mesons and baryons)
system, and also   to examine phase shifts (see Subsection 
\ref{subsec:scatteringanalysis}). The following $\pi$ and $N$ annihilation interpolators
are used to extract energies of the single hadrons with momenta ${\mathbf n}\,2\pi/ L$
(these are also used as building blocks for  interpolators in the Roper channel): 
  \begin{align}
  \label{pi}
 \pi^+(\mathbf{n})& =\sum_{\mathbf x} \bar d({\mathbf x},t)\gamma_5 u({\mathbf x},t) \mathrm{e}^{i\mathbf{x\cdot n}\frac{2\pi}{L}}  \\
   \pi^0(\mathbf{n}) &=\tfrac{1}{\sqrt{2}}\sum_{\mathbf x} [\bar d({\mathbf x},t)\gamma_5 d({\mathbf x},t)-\bar u({\mathbf x},t)\gamma_5 u({\mathbf x},t)] \mathrm{e}^{i\mathbf{x\cdot n}\frac{2\pi}{L}}\nonumber
 \end{align}
 and 
 \begin{align}
 \label{N}
 & N^{i}_{m_s=1/2}(\mathbf{n})= {\cal N}^{i}_{\mu=1}(\mathbf{n})\;,\   N^{i}_{m_s=-1/2}(\mathbf{n})= {\cal N}^{i}_{\mu=2}(\mathbf{n}) \\
 & {\cal N}^i_\mu(\mathbf{n})\!=\!\sum_{\mathbf{x}} \epsilon_{abc} [u^{aT}(\mathbf{x},t)  \Gamma_2^i d^b (\mathbf{x},t)] ~[\Gamma_1^i q^c(\mathbf{x},t)]_{\mu}~\mathrm{e}^{i\mathbf{x\cdot n}\frac{2\pi}{L}}\nonumber\\
 & i=1,2,3:\quad (\Gamma_1^i,\Gamma_2^i)=(\mathbf{1},C\gamma_5),~(\gamma_5,C),~(i\mathbf{1},C\gamma_t\gamma_4)\nonumber
  \end{align}
Three standard choices  for $\Gamma_{1,2}$ are used.  The 3rd quark is $q=u$ for the
proton and $q=d$ for the neutron.  Equation (\ref{N}) is in Dirac basis and the upper two
components ${\cal N}_{\mu=1,2}$ of the Dirac four spinor ${\cal N}_{\mu}$ are the ones
with positive parity at zero momentum.  The spin component $m_s$ in $N_{m_s}$ is a good
quantum number for $\mathbf{p}=0$ or  $\mathbf{p}\propto e_z$, which is employed to
determine energies  in Table \ref{tab:singleH}. It is not a good quantum number for
general $\mathbf{p}$ and it denotes the spin component $m_s$ of the corresponding field at
rest. The ``non-canonical" fields $N_{m_s}(\mathbf{n})$ (\ref{N}) built only from
upper-components have the desired transformation properties under rotation $R$ and
inversion $I$, which are necessary to build two-hadron operators \cite{Prelovsek:2016iyo}:
 \begin{align}
 \label{transf}
 RN_{m_s}(\mathbf{n})R^\dagger \!\!&=\!\!\! \sum_{m_s'} D^{1/2}_{m_sm_s'}(R^\dagger) N_{m_s'} (R\mathbf{n}), \nonumber\\
 R\pi(\mathbf{n})R^\dagger&=\pi(R\mathbf{n)}\nonumber\\
 I N_{m_s}(\mathbf{n})I &= N_{m_s}(-\mathbf{n}),\nonumber\\
  I\pi(\mathbf{n})I&=-\pi(-\mathbf{n})~.
 \end{align}  
Interpolators with narrower quark  sources are used for the determination of the masses
and energies of $\pi$ and $N$. Those are  collected in Table \ref{tab:singleH}, where they
are compared to energies $E^c$ expected in the continuum limit $a\to 0$.

\begin{table*}[t]
\begin{ruledtabular}
\begin{tabular}{c  c  c  c c c c c}
hadron & $\mathbf{n}=\tfrac{\mathbf{p}L}{2\pi} $  &interpol. & fit range & fit type & $\chi^2$/dof  & $E\,a$ (lat) &
 $E^{c}a=a \sqrt{m^2+\mathbf{p}^2}$  \\
\hline
$\pi$ & (0,0,0) &   $ \pi$  &  8-18 & cosh+exp, c  &   0.99& $0.07558 \pm 0.00098$  &  \\
$\pi$ & (0,0,1) &   $ \pi$  &  6-20  &  2 exp, c      &    1.91 &  $0.2049 \pm 0.0023 $ &0.2104\\
\hline
$N$ & (0,0,0) &      $N_n^{1,3}$   & 4-12  &  2 exp, c & 0.39&  $0.4455 \pm  0.0056$  &  \\
$N$ & (0,0,1) &      $N_n^{1,3}$   & 4-12  &  2 exp, c & 0.54 &  $0.4920 \pm 0.0072$   & 0.4864 \\
 \end{tabular}
\end{ruledtabular}
\caption{  The energies of single hadrons $\pi$ and $N$ for two relevant momenta, based on configuration set "all-4".  Energies in GeV are obtained by multiplying with $1/a\simeq 2.17~$GeV.  
}\label{tab:singleH}
\end{table*}
  
\subsection{Interpolating fields for the Roper channel}
 
Our central task is to calculate the energies of the eigenstates $E_n$ with $J^P=1/2^+$
and total momentum zero, including multi-particle states. We want to cover the energy
range up to approximately $1.65~$GeV, which is  relevant for the Roper region. The
operators with these quantum numbers have to be carefully constructed.   Although $qqq$
interpolators in principle couple also to multi-hadron intermediate states in dynamical
QCD, the multi-hadron eigenstates are often not established  in practice unless the
multi-hadron interpolators are also employed in the correlation matrix.

We apply 10  interpolators $O_{i=1,...,10}$ with   $P=+$,  $S=1/2$, $(I,I_3)=(1/2,1/2)$
and total momentum zero  \cite{Prelovsek:2016iyo} ($P$ and $m_s$ are good continuum
quantum numbers in this case). For $m_s=1/2$, we have 
      \begin{align}
     O_{1,2}^{N\pi}&=
    -\sqrt{\tfrac{1}{3}} ~\bigl[p^{1,2}_{-\frac{1}{2}}(-e_x) \pi^0(e_x)-p^{1,2}_{-\frac{1}{2}}(e_x) \pi^0(-e_x)\nonumber\\
    & \qquad \qquad-i p^{1,2}_{-\frac{1}{2}}(-e_y) \pi^0(e_y)+i p^{1,2}_{-\frac{1}{2}}(e_y) \pi^0(-e_y)\nonumber\\
     & \qquad \qquad+ p^{1,2}_{\frac{1}{2}}(-e_z) \pi^0(e_z)-p^{1,2}_{\frac{1}{2}}(e_z) \pi^0(-e_z)\bigr]\nonumber\\
      & \quad +\sqrt{\tfrac{2}{3}} ~\bigl[\{p\to n, \pi^0\to \pi^+\}\bigr] \quad [narrower] \nonumber\\
   O_{3,4,5}^{N_w}&=p^{1,2,3}_{\frac{1}{2}}(0)\quad [wider]\nonumber\\
    O_{6,7,8}^{N_n}&=p^{1,2,3}_{\frac{1}{2}}(0)\quad [narrower]\nonumber\\
   O_{9,10}^{N\sigma}&=p^{1,2}_{\frac{1}{2}}(0) \sigma(0) \quad [narrower] 
    \label{O}
\end{align} 
where these are the annihilation fields and   
\be
\label{sigma}
\sigma(0)=\tfrac{1}{\sqrt{2}}\sum_{\mathbf x} [\bar u({\mathbf x},t)u({\mathbf x},t)+\bar d({\mathbf x},t) d({\mathbf x},t)]~.
\ee   
The momenta of fields in units of $2\pi/L$ are given in parenthesis with $e_x$, $e_y$, and
$e_z$ denoting the unit vectors in $x, y$, and $z$ directions, while the lower index on
$N=p,n$ is $m_s$. All quarks have the same smearing width (narrower or wider in Fig.
\ref{fig:smearing}) within one interpolator. The $O^{N\pi}$ was constructed in
\cite{Prelovsek:2016iyo}, while  factors with square-root are Clebsch-Gordan coefficients
related to isospin. For $m_s=-1/2$, $p_{1/2}$ and $n_{1/2}$ gets replaced by $p_{-1/2}$
and $n_{-1/2}$ in $O_{3-10}$, while $O_{1,2}$ becomes  \cite{Prelovsek:2016iyo} 
\begin{align}
     O_{1,2}^{N\pi}&=
    -\sqrt{\tfrac{1}{3}} ~\bigl[p^{1,2}_{\frac{1}{2}}(-e_x) \pi^0(e_x)-p^{1,2}_{\frac{1}{2}}(e_x) \pi^0(-e_x)\nonumber\\
    & \qquad \qquad+i p^{1,2}_{\frac{1}{2}}(-e_y) \pi^0(e_y)-i p^{1,2}_{\frac{1}{2}}(e_y) \pi^0(-e_y)\nonumber\\
     & \qquad \qquad- p^{1,2}_{-\frac{1}{2}}(-e_z) \pi^0(e_z)+p^{1,2}_{-\frac{1}{2}}(e_z) \pi^0(-e_z)\bigr]\nonumber\\
      & \quad +\sqrt{\tfrac{2}{3}} ~\bigl[\{p\to n, \pi^0\to \pi^+\}\bigr] \quad [narrower] 
\label{Osd}
\end{align}

 The basis (\ref{O})  contains  conventional $qqq$ fields as well as the most relevant multi-hadron 
components.  The non-interacting levels below $1.65~$GeV  are  $N(0)$, $N(1)\pi(-1)$, 
$N(0)\pi(0)\pi(0)$ and, assuming zero width approximation, $N(0) \sigma(0)$. The      $N(2)\pi(-2)$, 
$N(1)\pi(-1)\pi(0)$ and others   are at higher energies. Here $O^{N\pi}$ corresponds to  $N(1)\pi(-1)$ in $p$-wave \cite{Prelovsek:2016iyo}. Our notation  implies projection to $J^P=\frac{1}{2}^+$ for all 
operators (e.g., $N(1)\sigma(-1)$ actually refers to $\sum_{\mu=1}^3 N(e_\mu)\sigma(-e_\mu)$).  
Interpolators $N(n)\pi(-n)$ with $n\geq 2$ are not incorporated, so we do not expect to find  those in the 
spectrum.  We implement only one type of $\sigma$ interpolator (\ref{sigma}) in $O^{N\sigma}$ and we  
expect that this represents a possible superposition of $N\pi\pi$ and $N\sigma$.\footnote{The $\sigma$ 
channel itself  was recently simulated with a number of interpolators in   \cite{Briceno:2016mjc}.}

On the discrete lattice the continuum rotation symmetry group is reduced to the discrete
lattice double-cover group $O_h^2$. The states with the continuum quantum number
$J^P=1/2^+$ transform according to the  $G_1^+$ irreducible representation  on the
lattice. All operators (\ref{O})  indeed transforms according to  $G_1^+$
 \begin{align}
 RO^{m_s}_i(0)R^\dagger &= \sum_{m_s'} D^{1/2}_{m_sm_s'}(R^\dagger) O^{m_s'}_i (0),\ \nonumber \\
 I O^{m_s}_i(0)I & = O_i^{m_s}(0),
  \end{align}
as can be checked by using the transformations of individual fields $N$, $\pi$, $\sigma$
(eqn. \ref{pi}, \ref{N}, \ref{sigma}). The $N\pi$ operator  with such transformation
properties was constructed using the projection, partial-wave and helicity methods  
\cite{Prelovsek:2016iyo}, all leading to $O_{1,2}^{N\pi}$  in eqns. (\ref{O},\ref{Osd}).  
The partial-wave method indicates that it describes $N\pi$ in $p$-wave.

We restrict our calculations to zero total momentum since parity is a good quantum number
in this case. The positive parity states with $J=1/2$ as well as  $J\geq 7/2$ appear in  
the relevant irreducible representation $G_1^+$ of $O_h^2$.  The observed baryons with 
$J\geq 7/2$   lie above $1.9~$GeV, therefore  this does not present a complication  for
the energy region of our interest. We do not consider the system with non-zero total
momenta since $1/2^+$ as well as $1/2^-$ (and others) appear in the same irreducible
representation \cite{Gockeler:2012yj}, which would be a significant complication
especially due to the negative parity states $N(1535)$ and $N(1650)$. 
  
  \subsection{Wick contractions for the Roper channel} 
  
The $10\times 10$ correlation function $C_{ij}(t)$ (\ref{C}) for the Roper channel is
obtained after evaluating the Wick contractions for any pair of source $\bar{O}_j $ and
sink $O_i$.  The number of Wick contractions involved in computing the correlation
functions between our interpolators (eqn. \ref{O}) are tabulated in Table \ref{tb:Wick}.
  
\begin{table}[h!]  \begin{tabular}{ c| c c c } 
   $O_i\backslash O_j$ & $O^N$ & $O^{N\pi}$ & $O^{N\sigma}$ \\
   \hline
   $O^N$ & 2   & 4   & 7 \\
   $O^{N\pi}$ & 4  &  19 &   19\\
   $O^{N\sigma}$ & 7   & 19  &  33 
  \end{tabular}  
\caption{Number of Wick contractions involved in computing correlation functions between
interpolators in eqn. (\ref{O}).}
\label{tb:Wick}
\end{table}     
  \vspace{0.2cm}
  
The $O^N\leftrightarrow O^N$ contractions have been widely used in the past.
\footnote{Footnote added after publication: the $N\pi$ contribution to correlators $O^N\leftrightarrow O^N$ 
with local operators has been determined via ChPT in \cite{Bar:2015zwa}.} The  19
Wick-contractions $O^{N\pi}\leftrightarrow O^{N\pi}$ and 4 Wick contractions
$O^{N}\leftrightarrow O^{N\pi}$ are  the same  as  in the Appendix of  \cite{Lang:2012db},
where the negative-parity  channel was studied.  The inclusion of $O^{N\sigma}$ introduces
additional $2\cdot 7 + 2\cdot 19+33$ Wick contractions, while the inclusion of three
hadron interpolators like  $N\pi\pi$ would require many more. We evaluate all necessary
contractions  in Table \ref{tb:Wick} using the distillation method \cite{Peardon:2009gh}
discussed in Section \ref{sec:Dis}.

Appendix \ref{sec:wick} illustrates how to handle the   spin components in evaluating
$C(t)$, where one example of the Wick contraction $\langle\Omega|O^{N\pi}\bar
O^{N}|\Omega\rangle$ is considered. 
  
 \begin{figure*}[!htb]
\begin{center} 
\includegraphics*[height=70mm,clip]{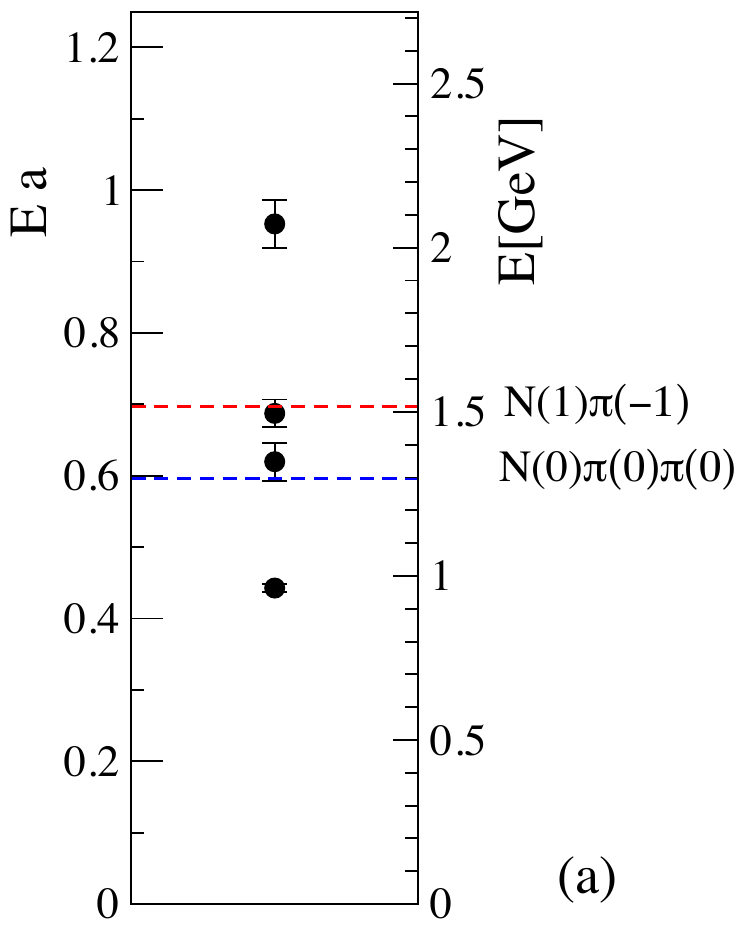}$\qquad$
\includegraphics*[height=78mm,clip]{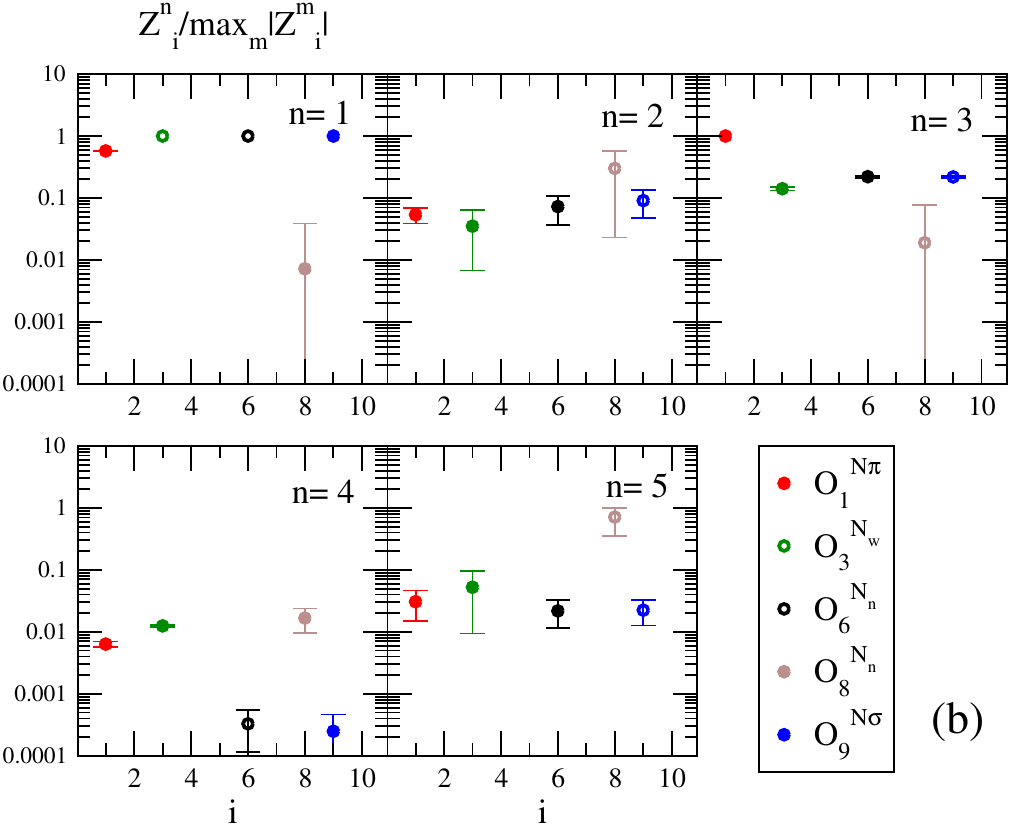}
\end{center}
\caption{  The  eigenenergies $E_n$ (a) and normalized overlaps $Z_i^n=\langle \Omega| O_i|n\rangle$ (b), which result from correlation matrix (\ref{C}) based on the complete interpolator set (\ref{O_complete}).  Left pane (a): The energies $E_n$ from lowest ($n=1$) to highest ($n=4$). The horizontal dashed lines  represent  the energies $m_N+2m_\pi$ and $E_{N(1)}+E_{\pi(-1)}$ of the expected  multi-hadron states  in the non-interacting limit.   Right pane (b):  the  ratios of overlaps $Z_i^n$ with respect to  the largest among $|Z_i^{m=1,...5}|$; these     ratios   are independent on the normalization of $O_i$. The full and empty symbols correspond to the positive  and negative $Z_i^n$, respectively   ($Z_i^n$ are  almost real).  Configuration set "all-4" is used.   }
\label{fig:E_final}
\end{figure*}

\begin{figure}[!htb]
\begin{center}
\includegraphics*[width=0.35\textwidth,clip]{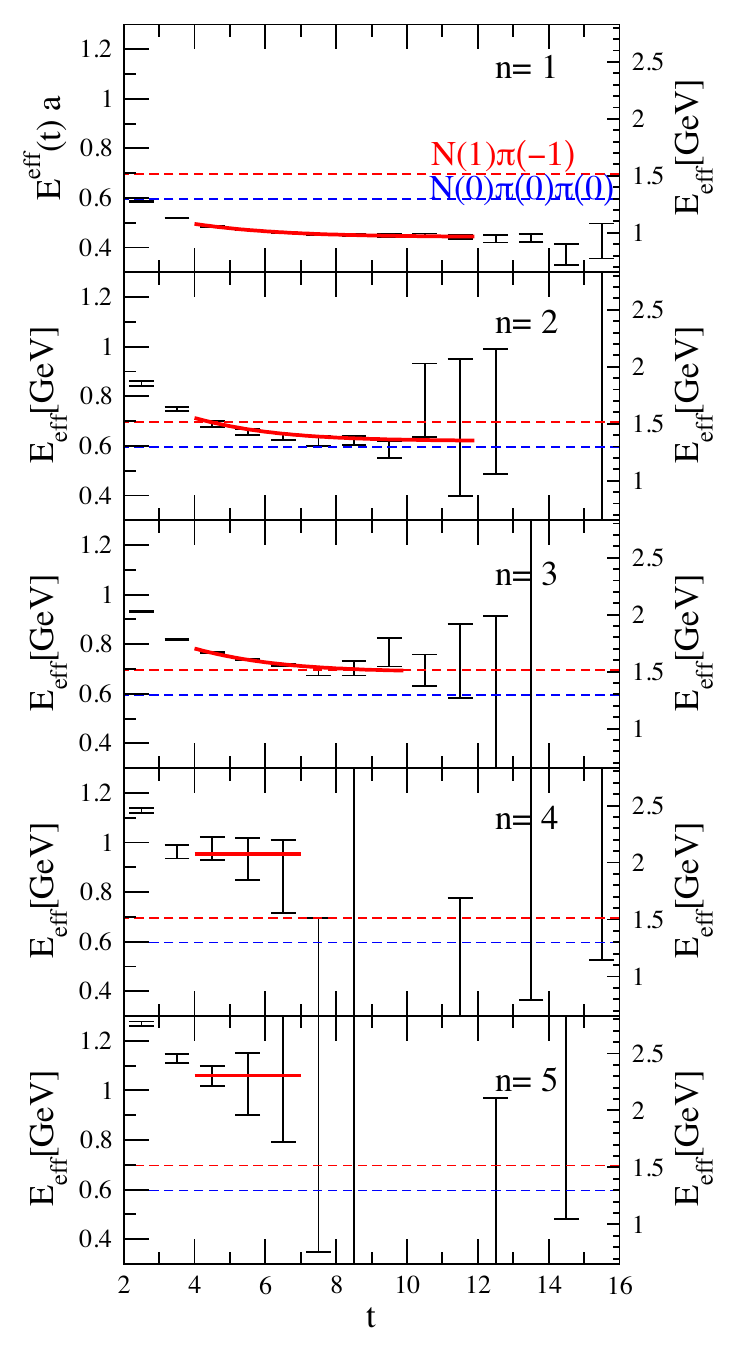}
\end{center}
\caption{  The  effective energies   $E_n^{eff}(t)=\log[\lambda^{(n)}(t)/\lambda^{(n)}(t+1)]\to E_n$ of eigenvalues $\lambda^{(n)}$.   These correspond to  the energies of eigenstates   $E_n$ in Fig. \ref{fig:E_final}a and Table \ref{tab:E_final}. It is based on the complete interpolator set (\ref{O_complete}) and configuration set "all-4". The fits of $\lambda^{(n)}(t)$ that render $E_n$ are also presented.  Non-interacting energies of   $N(0)\pi(0)\pi(0)$ and $N(1)\pi(-1)$ are shown with dashed lines.    }
\label{fig:Eeff_final}
\end{figure}

\begin{table}[t]
\begin{ruledtabular}
\begin{tabular}{c  c  c  c c  }
eigenstate  & fit  & fit & $\chi^2$/dof  & $E\,a$\\
 $n$ &   range & type &   &     \\
\hline 
1 &4-12 & 2 exp, c&0.50 & $0.4427\pm 0.0055$ \\
2 & 4-12& 2 exp, c&1.04 & $0.6196\pm 0.0266$ \\
3 &4-10 & 2 exp, c&0.88 & $0.6873\pm 0.0195$ \\
4 &4-7   &1 exp, c  &0.32 & $0.9527 \pm 0.0338$ \\
 \end{tabular}
\end{ruledtabular}
\caption{ The final energies $E_n$ of eigenstates in the Roper channel, which correspond to  
Fig. \ref{fig:E_final}a and effective masses in Fig. \ref{fig:Eeff_final}. 
They are obtained from correlated fits based on complete interpolator 
set (eqn. \ref{O_complete}) and configuration set ''all-4". Energies 
in GeV can be obtained by multiplying with $1/a\simeq 2.17~$GeV. }\label{tab:E_final}
\end{table}

\begin{figure}[!htb]
\begin{center} 
\includegraphics*[height=65mm,clip]{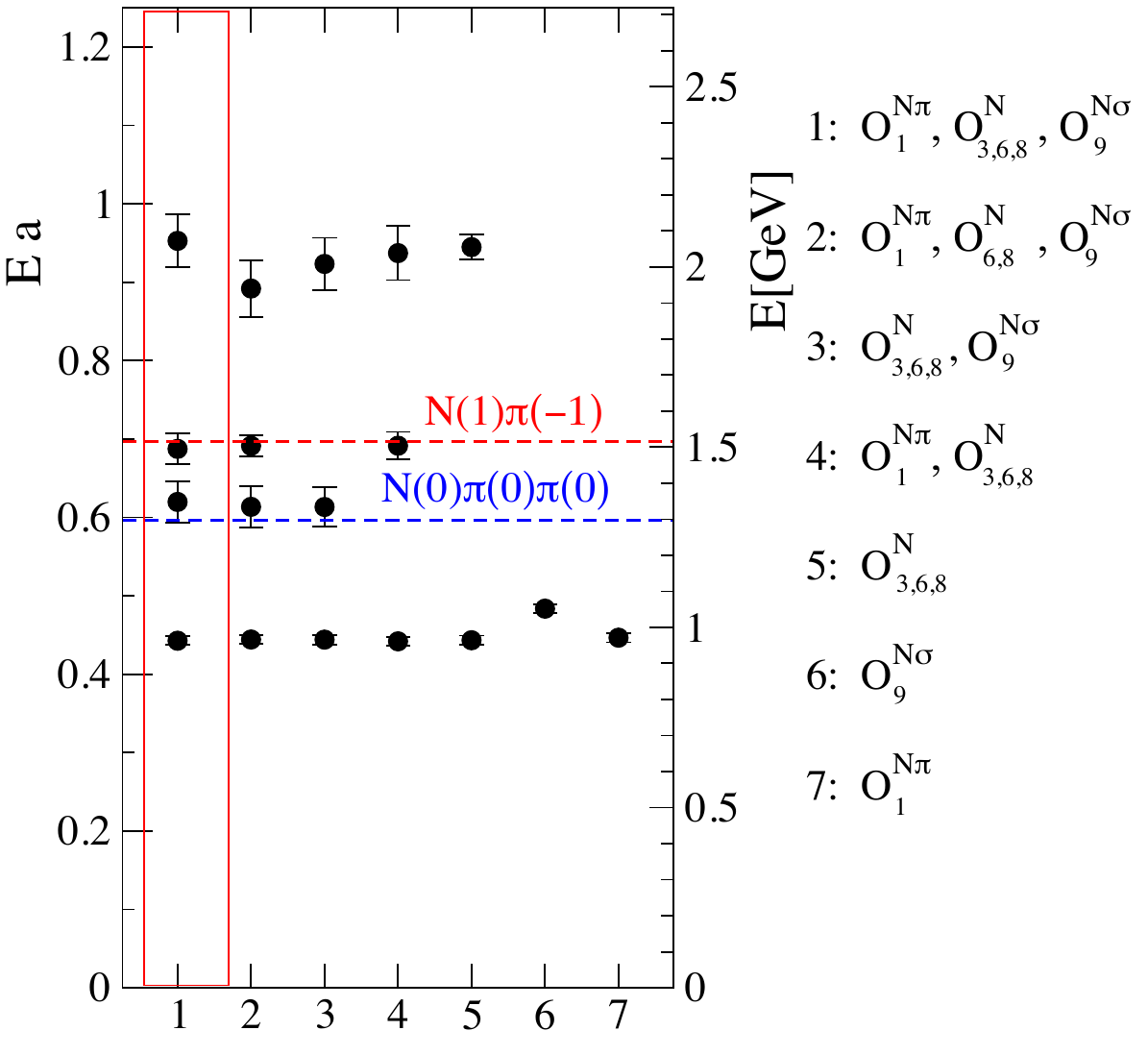} 
\end{center}
\caption{The energies of eigenstates  $E_n$ for various choices of interpolator basis
(\ref{O}) used in correlation matrix (\ref{C},\ref{gevp}). The reference choice 1
representing the complete interpolator set    $O_1^{N\pi},~
O^{N_w}_3,~O^{N_n}_{6,8},~O_{9}^{N\sigma}$  (\ref{O_complete}) is  highlighted. One or
more interpolators are removed for other choices.  The horizontal lines present
non-interacting energies of   $N(0)\pi(0)\pi(0)$ and $N(1)\pi(-1)$. Results are based on
configuration set "all-4".}
\label{fig:E_interp_set}
\end{figure} 

  \begin{figure}[!htb]
\begin{center} 
\includegraphics*[height=65truemm,clip]{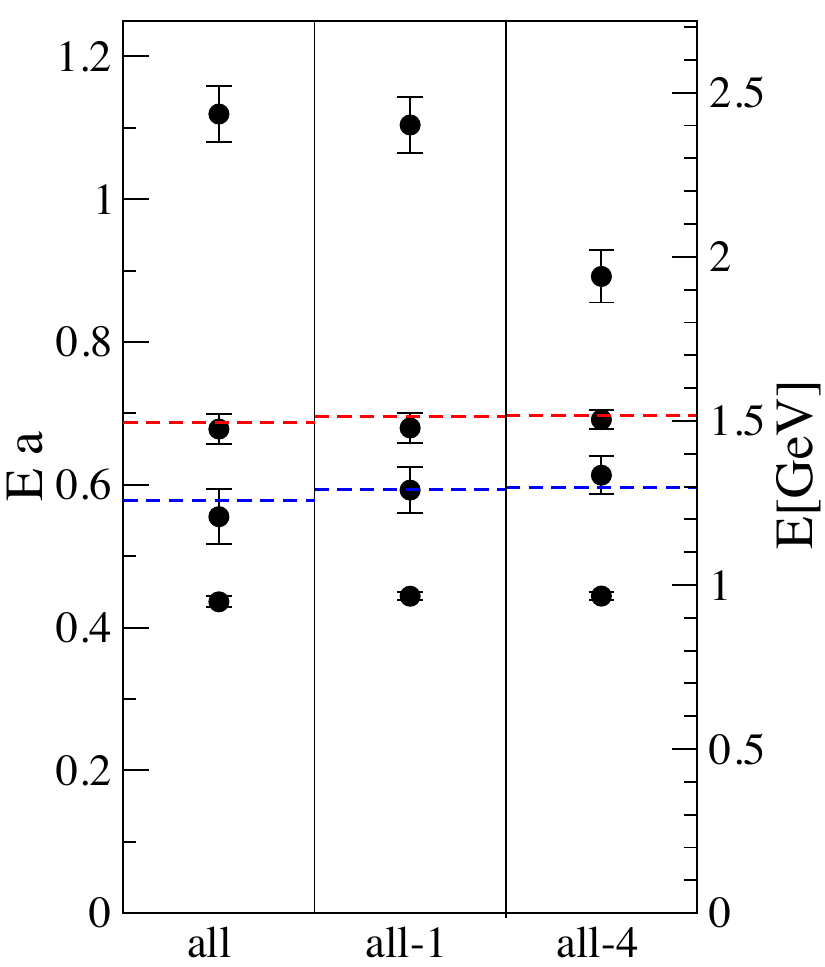} 
\end{center}
\caption{The energies $E_n$ are determined on  all 197 configurations ("all"),  on 196
configurations ("all-1"), and on 193 configurations ("all-4"), as described in Section
\ref{sec:conf}. The values are based on the interpolator set
$O_1^{N\pi},~O^{N_n}_{6,8},~O_{9}^{N\sigma}$ which gives smaller statistical errors than
set (\ref{O_complete}) for "all" and "all-1". The horizontal lines present non-interacting
energies of   $N(0)\pi(0)\pi(0)$ and $N(1)\pi(-1)$ for the corresponding configuration
sets. }
\label{fig:E_config_set}
\end{figure}

\section{Results} \label{sec:results}

 \subsection{Energies and overlaps}

Our main result  are the energies of the eigenstates in  the $J^P=1/2^+$ channel, shown 
in Fig. \ref{fig:E_final}a. These  are based on the $5\times 5$ correlation matrix
(\ref{C}) for the subset of  interpolators  (\ref{O})
\be
\label{O_complete}
\mathrm{complete\ interpolator\ set:}\ O_1^{N\pi},~ O^{N_n}_3,~O^{N_w}_{6,8},~O_{9}^{N\sigma}\;,
\ee
which we refer to as the "complete set" since it contains all types of interpolators. 
Adding other interpolators to this basis, notably $ O_{2,4,7,10}$, which include the
$N^{i=2}$ interpolator\footnote{It has been observed already earlier, e.g. 
\cite{Brommel:2003jm}, that this interpolator shows no plateau behavior in the effective
energy.}, makes the eigenenergies noisier. The eigenenergies $E_n$   are obtained  from
the   fits of the eigenvalues $\lambda^{(n)}(t)$ (\ref{gevp}),  with fit details  in Table
\ref{tab:E_final}. The horizontal dashed lines represent  the energies of the expected 
multi-hadron states $m_N+2m_\pi$ and $E_{N(1)}+E_{\pi(-1)}$ in the non-interacting limit
(the  individual hadron energies measured on our lattice and given in Table \ref{tab:singleH}
are used for this purpose throughout this work). The study of this channel with  
almost physical pion mass is  challenging as far as statistical errors are concerned. 
This can be seen from the effective energies  in Fig. \ref{fig:Eeff_final} which give
eigenenergies   in the plateau region.

The ground state ($n=1$) in Fig. \ref{fig:E_final}a represents the nucleon. The 
first-excited eigenstate ($n=2$) lies near $m_N+2m_\pi$ and appears to be close to
$N(0)\pi(0)\pi(0)$ in the non-interacting limit. The next eigenstate $n=3$   lies near the
non-interacting energy $E_{N(1)}+E_{\pi(-1)}$. It dominantly couples to $O^{N\pi}$ and  we
relate it to     $N(1)\pi(-1)$ in the non-interacting limit. Further support in favor of
this identification for levels $n=2,3$ will be given in the discussion of Figs.
\ref{fig:E_interp_set} and \ref{fig:E_config_set}. The most striking feature of the
spectrum is that there are only three eigenstates below $1.65~$GeV, while the  other
eigenstates appear at higher energy.

The overlaps of these eigenstates with various operators are presented in Fig.
\ref{fig:E_final}b.    The nucleon ground state $n=1$   couples well with all
interpolators that contain $N^1$.     The operator $O^{N\pi}$ couples well with eigenstate
$n=3$, which gives further support that this state is related to $N(1)\pi(-1)$.   The
operator $O^{N\sigma}$ couples best with the nucleon ground state, which is not surprising
due to the presence of the   Wick contraction  where the isosinglet $\sigma$  
(\ref{sigma})  annihilates and the remaining $N^1$  couples to the nucleon. Interestingly,
the $O^{N\sigma}$  has similar couplings to the eigenstates $n=2$ and $n=3$, which are
related to $N(0)\pi(0)\pi(0)$ and $N(1)\pi(-1)$ in the non-interacting limit.  One would
expect $|\langle \Omega| O^{N\sigma} |n=2\rangle| \gg  |\langle \Omega|O^{N\sigma}
|n=3\rangle|$ if the channel $N\pi$ were decoupled from $N\sigma/N\pi\pi$. Our overlaps 
$Z_{i=9}^{n=2,3}$  suggest that the channels are significantly coupled.  The scenario
where  the coupled-channel scattering might be crucial for the Roper resonance will
discussed in Section \ref{sec:discussion}.

The features of the spectrum for various choices of the interpolator basis are
investigated in Fig. \ref{fig:E_interp_set}. The complete set (\ref{O_complete}) with all
types of interpolators is highlighted as choice 1.  If the operator $O^{N\pi}$ is removed
(choice 3) the eigenstate with energy $\simeq E_{N(1)}+E_{\pi(1)}$ disappears, so the
$N\pi$ Fock component is important for this eigenstate. The eigenstate with energy $\simeq
m_N+2m_\pi$ disappears  if $O^{N\sigma}$ is removed (choice 4), which suggests that this
eigenstate is  dominated by  $N(0)\pi(0)\pi(0)$, possibly mixed with $N(0)\sigma(0)$. Any interpolator individually
renders the nucleon as a ground state (choices 5,6,7).

All previous lattice simulations, except for \cite{Kiratidis:2016hda}, used just $qqq$
interpolators. This  is represented by the choice 5, which renders the nucleon, while the
next state is  above $1.65~$GeV; this result is in agreement with most of the previous
lattice results based on $qqq$ operators, discussed in the Introduction.  No
interpolator basis renders  more than three eigenstates below $1.65~$GeV.

The most striking feature of the spectra in Figs. \ref{fig:E_final} and
\ref{fig:E_interp_set} is the absence of any additional eigenstate  in the energy region
where the Roper resonance resides in experiment. The eigenstates $n=2,3$ lie in this
energy region, but two eigenstates related to $N(0)\pi(0)\pi(0)$ and $N(1)\pi(-1)$ are
inevitably expected there in dynamical QCD, even in absence of the interactions between
hadrons.

A further indication that eigenstate $n=2$ is dominated by $N(0)\pi(0)\pi(0)$ is presented
in Fig. \ref{fig:E_config_set}, where the spectrum from all configurations is compared to
the spectrum based on configuration sets "all-4" (shown in other figures) and "all-1".
The horizontal dashed lines indicate non-interacting energies obtained from the
corresponding sets. Only the central value of   $E_{2}$ and  $m_N+2m_\pi$ visibly depend
on the configuration set. The variation of $m_N+2m_\pi$ is due to the variations of
$m_\pi$ pointed out in Section \ref{sec:conf}. The eigenstate $n=2$ appears to track  the
threshold $m_N+2m_\pi$, which  suggests that its Fock component $N(0)\pi(0)\pi(0)$ is
important. Note that the full configuration set gives larger statistical errors, as
illustrated via effective masses in  Fig.  \ref{fig:Eeff_set}   of Appendix
\ref{app:effmasses}.  

\subsection{Scattering phase shift}\label{subsec:scatteringanalysis}
        
In order to discuss the $N\pi$ phase shift, we consider the elastic approximation where
$N\pi$ scattering is decoupled from the $N\pi\pi$ channel. In this case, the $N\pi$ phase
shift  $\delta$ can  be determined from the eigenenergy $E$  of the interacting state
$N\pi$ via 
  L\"uscher's relation \cite{Luscher:1990ux,Luscher:1991cf} 
  \be
  \label{luscher}
   \delta(p)=\mathrm{atan}\biggl[\frac{\sqrt{\pi} p L}{2\,Z_{00}(1;(\tfrac{pL}{2\pi})^2)}\biggr],\  E=E_{N(p)}+ E_{\pi(p)}
  \ee   
where $E_{H(p)}=\sqrt{m_H^2+p^2}$  applies in the continuum limit. The eigenenergy $E$ 
($E_3$ from basis $O^{N\pi,N,N\sigma}$ or  $E_2$ from  $O^{N\pi,N}$)   has sizable error
for this ensemble with close-to-physical pion mass. It lies close to the non-interacting
energy $E_{N(1)}+E_{\pi(1)}$, as can be seen in Figs. \ref{fig:E_final},
\ref{fig:Eeff_final} and \ref{fig:Eeff_set}. We find  that the resulting energy shift
$\Delta E=E-E_{N(1)}-E_{\pi(1)}$ is consistent with zero (modulo $\pi$) within the errors.
This implies that the phase shift $\delta$ is zero within a large  statistical error.

We verified this using a number of   choices to extract $\Delta E$ and $\delta$. The
interpolator set $O^{N\pi,N}$ rightmost column of Fig. \ref{fig:Eeff_set}) that imitates
the elastic $N\pi$ scattering served as a main choice, while it was compared to other sets
also.  Correlated and uncorrelated fits of $E$ as well as $E_{N(1)}+E_{\pi(1)}$ were
explored  for various fit-ranges. Further choices of dispersion relations $E_\pi(p)$ and
$E_N(p)$ that match lattice energies at $p=0,1$ in Table \ref{tab:singleH} (e.g.,
interpolation of $E^2$ linear in $p^2$)  were  investigated within the L\"uscher analysis
to arrive at same conclusions.

\begin{figure}[!tb]
\begin{center}
\includegraphics*[width=0.45\textwidth,clip]{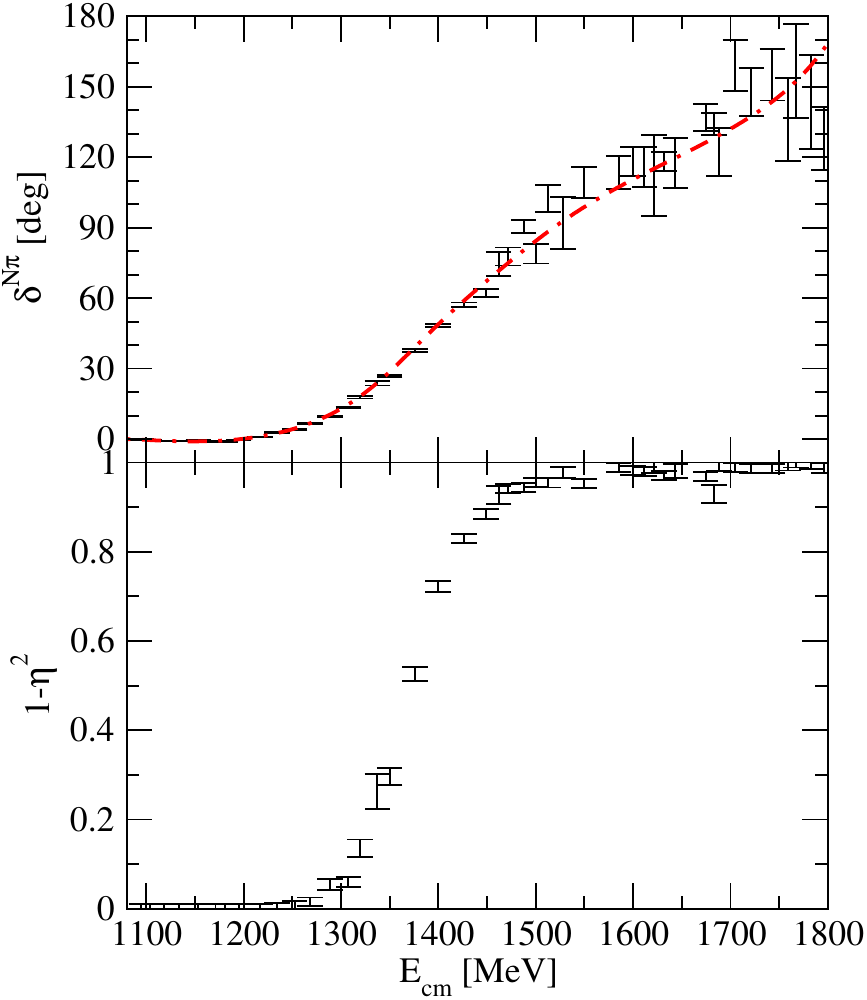} 
\end{center}
\caption{  The experimental phase shift $\delta$ and  inelasticity $1-\eta^2$ as extracted 
by the GWU group \cite{Workman:2012hx} (solution WI08). The  dot-dashed line is a smooth 
interpolation that is used in Section \ref{sec:discussion_elastic}.}
\label{fig:P11_exp}
\end{figure}    

\section{Discussion and interpretation} \label{sec:discussion}
        
Here we discuss the implications of our results, in particular  that  only three
eigenstates are found below $1.65$ GeV. These appear  to be associated with
$N(0),~N(0)\pi(0)\pi(0)$ and $N(1)\pi(-1)$ in the non-interacting limit.

The experimental $N\pi$ scattering data for the amplitude $T=(\eta e^{2i\delta}-1)/(2i)$
for this  ($P_{11}$) channel  are  shown in Fig. \ref{fig:P11_exp} \cite{Workman:2012hx}\footnote{ The experimental data comes from the GWU homepage {\tt
gwdac.phys.gwu.edu}}. The channel  is complicated by the fact that $N\pi$ scattering is
not elastic above the $N\pi\pi$ threshold  and  the inelasticity is sizable already in the
energy region of the Roper resonance.

The  presence of the $N\pi\pi$ channel prevents rigorous investigation on lattice at the
moment. While the three-body channels have been treated analytically, see for example
\cite{Hansen:2015zga,Hansen:2016ync}, the scattering parameters have not been determined
in any channel within lattice QCD up to now. For this reason we consider implications  for
the lattice spectrum based on various simplified scenarios. By comparing our lattice
spectra to the predictions of these scenarios,  certain conclusions on the Roper resonance
are drawn.  

\begin{figure}[!htb]
\begin{center}  
\includegraphics*[width=0.45\textwidth,clip]{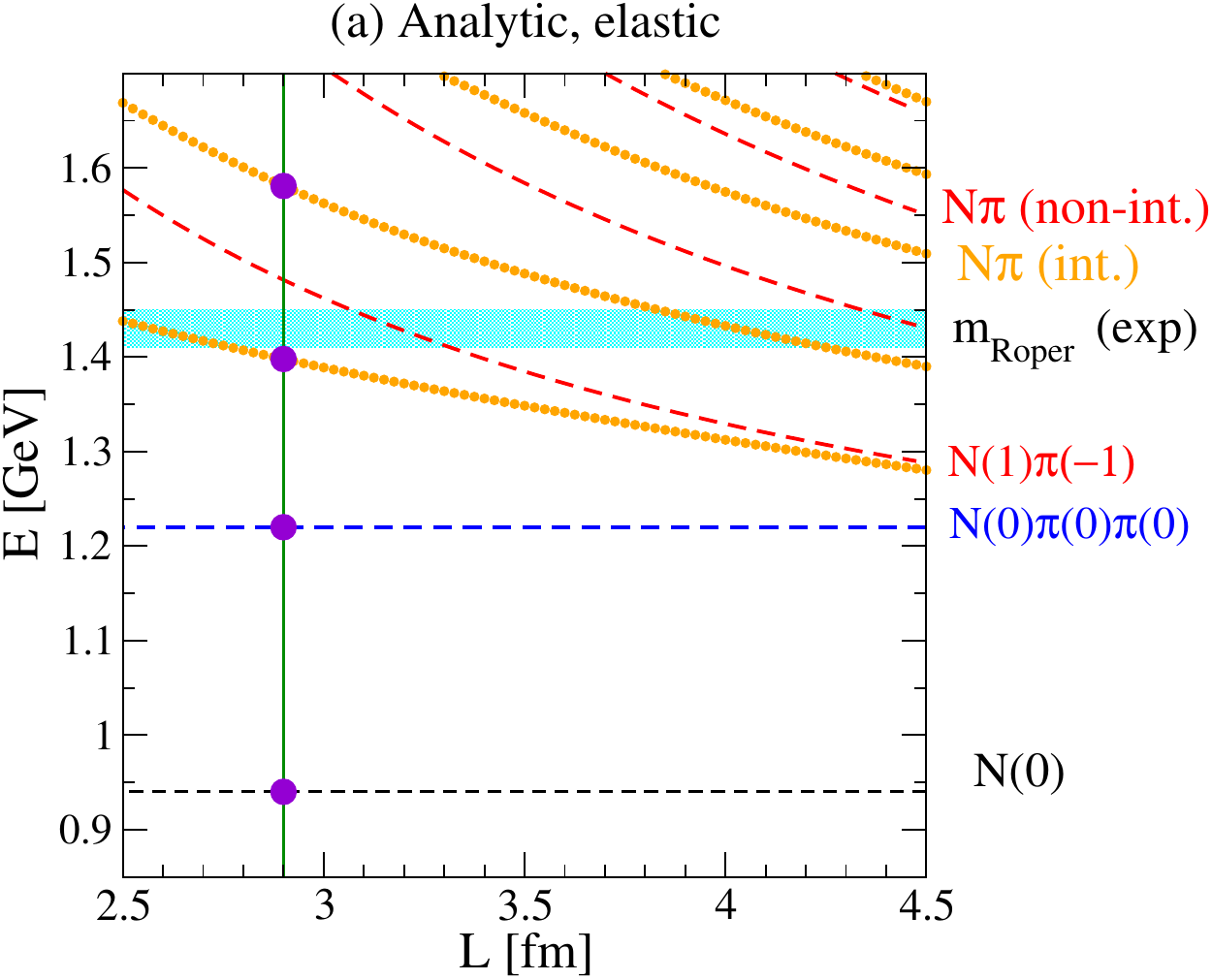}\\~\\
\includegraphics*[width=0.45\textwidth,clip]{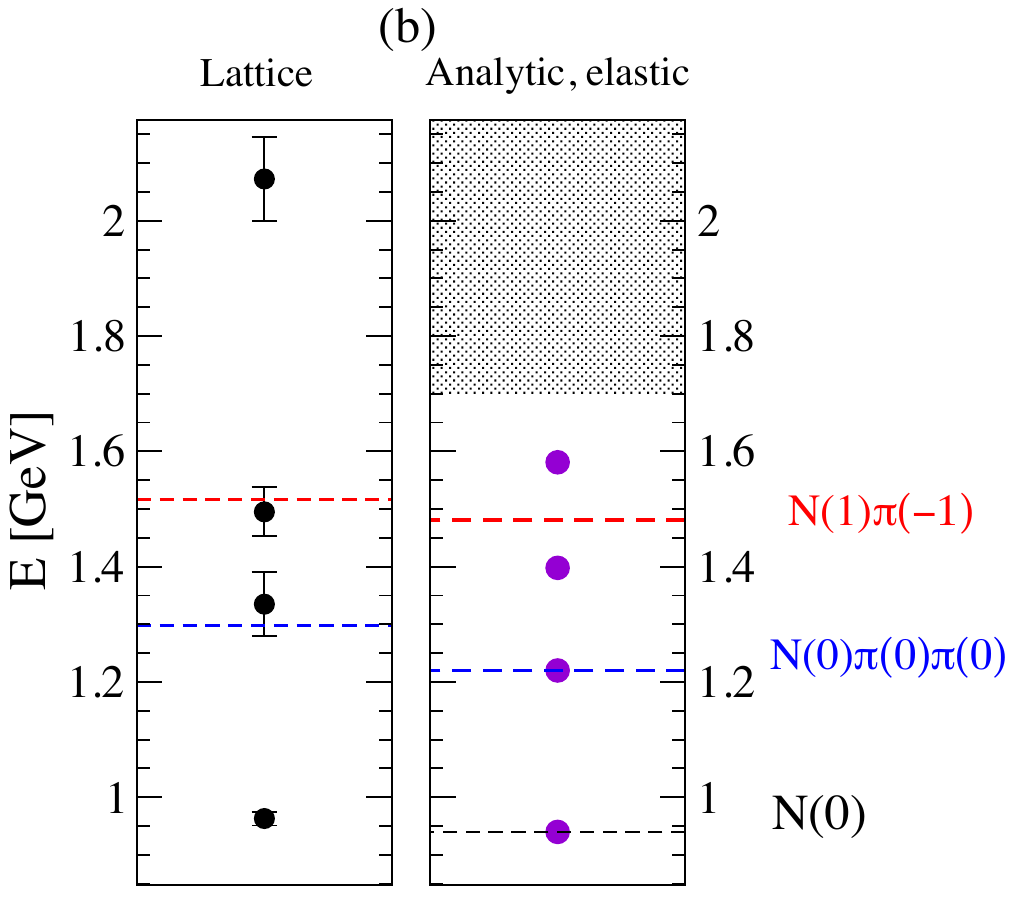}
\end{center}
\caption{ (a) Analytic prediction for the eigenenergies $E$ as a function of the lattice
size $L$, according to  (\ref{luscher}). The $N\pi$  and $N\pi\pi$   are assumed to be
decoupled, and $N\pi\pi$ is non-interacting. The curves show: non-interacting $N\pi$  (red
dashed), interacting  $N\pi$ based on experimental phase shift \cite{Workman:2012hx}  
(orange dotted), $N\pi\pi$ threshold (blue dashed), proton mass (black),  Roper mass (cyan
band). The experimental masses of hadrons are used.   (b) Left: energy values from our
simulation. (b) Right: the full violet circles show the analytic predictions for the
energies at our $L=2.9~$fm based on the experimental phase shift data and elastic
approximation (same as violet circles in upper pane).  We show only the energy region 
$E<1.7~$GeV where we aim to extract the complete spectrum  (there are additional 
multi-hadron states in the shaded region and we did not incorporate interpolator fields
for those).   
  }
\label{fig:E_analytic}
\end{figure}

 \subsection{\texorpdfstring{$N\pi$ scattering in elastic approximation}{}}   \label{sec:discussion_elastic}
        
 Let us examine what would be the lattice spectrum assuming experimental $N\pi$  phase
 shift  in the approximation when $N\pi$ is decoupled from  the $N\pi\pi$ channel. In
 addition we  consider no interactions in the $N\pi\pi$ channel. The elastic phase shift
 $\delta$ in Figure \ref{fig:P11_exp} allows to obtain the discrete energies $E$ as
 function of the spatial lattice size $L$ via L\"uscher's equation (\ref{luscher}) .

 Figure \ref{fig:E_analytic}a shows the non-interacting  levels for $N(0)$ (black),
 $N(0)\pi(0)\pi(0)$ (blue), and $N(1)\pi(-1)$ (red). These  are shifted by the interaction. Also
 plotted are the eigenstates (orange) in the interacting $N\pi$ channel derived from the
 experimental elastic phase shift with help of eqn. (\ref{luscher}). The elastic scenario
 should therefore render four eigenstates below 1.65 GeV at our $L\simeq 2.9~$fm,
 indicated by the violet circles in Figures  \ref{fig:E_analytic}a and
 \ref{fig:E_analytic}b.  Three non-interacting levels\footnote{These are three intercepts
 of dashed curves with vertical green line at $L=2.9~$fm.} below $1.65~$GeV  turn into
 four interacting levels (violet circles) at $L\simeq 2.9~$fm.  The Roper resonance phase
 shift passing $\pi/2$ is responsible    for the extra level.

Our actual lattice data   features only three eigenstates below $1.65~$GeV, and no extra
low-lying  eigenstate is found. Comparison in Figure \ref{fig:E_analytic}b indicates that
the lattice data  is qualitatively different from  the prediction of the resonating $N\pi$
phase shift for the low-lying Roper resonance, assuming it is decoupled from $N\pi\pi$.

 \subsection{\texorpdfstring{Scenarios with  coupled $N\pi-N\sigma-\Delta\pi$ scattering}{}}   
  
\begin{figure}[!htb]
\begin{center} 
\includegraphics*[width=0.49\textwidth,clip]{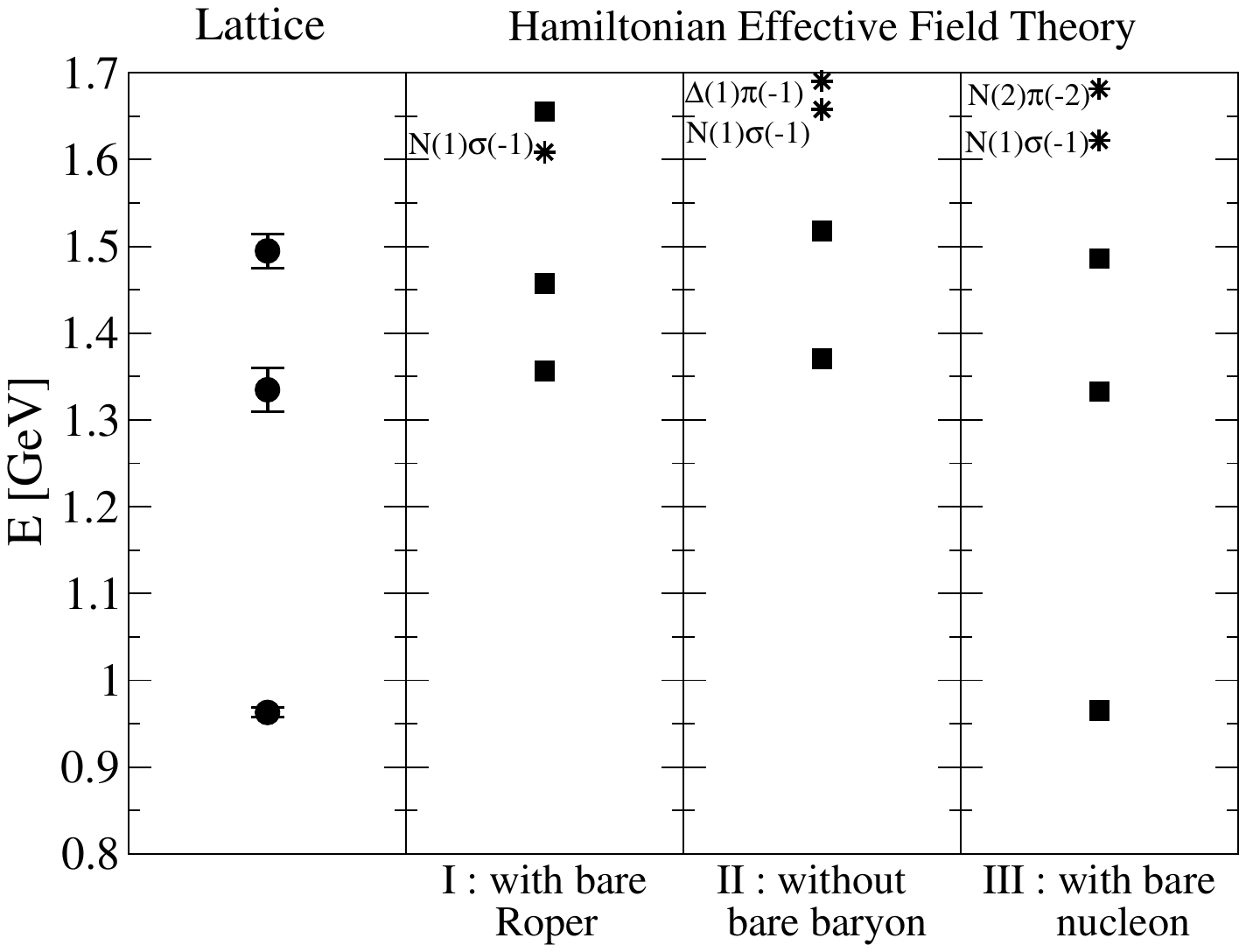}
\end{center}
\caption{ Analytic predictions for the lattice spectra at $m_\pi=156~$MeV and $L=2.9~$fm from the Hamiltonian Effective Field theory. These are  based on three scenarios concerning the Roper resonance  \cite{Liu:2016uzk}. Our lattice spectrum is shown with circles on the left.  Qualitative comparison   between the energies represented by  squares  and circles can be made, as discussed in the main text. }
\label{fig:HEFT}
\end{figure}

Our analysis does not show the resonance related level. One reason could be that the Roper
resonance is a truly coupled channel phenomenon and one has to include further
interpolators like $\Delta \pi$, N$\rho$ and an explicit $N\pi\pi$ three hadron
interpolator. The  scattering of  $N\pi-N\sigma-\Delta\pi$ in the Roper channel   was 
studied recently using Hamiltonian Effective Field Theory (HEFT) \cite{Liu:2016uzk}. The
$\sigma$ and $\Delta$ were assumed to be stable under the strong decay, which is a
(possibly serious) simplification. The free parameters were always fit to the experimental
$N\pi$ phase shift and describe the data well. Three models were discussed:
\begin{enumerate}
\item[I]
 The three channels are coupled with a low-lying bare Roper operator of type $qqq$. 
 \item[II] 
No bare baryon; the $N\pi$ phase shift is reproduced solely via coupled channels. 
 \item[III]
 The three channels are coupled only to a bare nucleon.  
 \end{enumerate}
The resulting Hamiltonian was  considered  in a finite volume leading to discrete
eigenenergies for  all three cases, plotted in Fig. \ref{fig:HEFT} for our parameters
$L=2.9~$fm and $m_\pi=156~$MeV \cite{Liu:2016uzk}.

In Fig. \ref{fig:HEFT} we compare our lattice spectra with the prediction for energies of $J^P=1/2^+$ states in three scenarios. The stars mark the
high-lying eigenstates $N(1)\sigma(-1)$, $\Delta(1)\pi(-1)$ and $N(2)\pi(-2)$
\cite{Liu:2016uzk}, which are not expected to be found in our study since we did not incorporate corresponding interpolators in
(\ref{O}). The squares denote predictions from the three scenarios that can be
qualitatively compared with our lattice spectra. 

Our lattice levels below $1.7~$GeV   disagree with  model  I based on bare Roper $qqq$
core, but are consistent with II and (preferred) III with no bare Roper $qqq$  core.  In
those scenarios the Roper resonance   is dynamically generated from the 
$N\pi/N\sigma/\Delta\pi$ channels, coupled also to a bare nucleon core in case III.   
Preference for  interpretations  II,III was reached also in  other phenomenological
studies \cite{Krehl:1999km,Schutz:1998jx,Liu:2016uzk,Matsuyama:2006rp}  and on the lattice
\cite{Kiratidis:2016hda}, for example.

\subsection{Hybrid baryon scenario}
        
Several authors, for example \cite{Golowich:1982kx,Kisslinger:1995yw}, have proposed that
the Roper resonance might be a hybrid baryon $qqqG$ with excited gluon field. This
scenario predicts   the longitudinal helicity amplitude $S_{1/2}$ to vanish
\cite{Li:1991yba},  which is   not supported by the measurement   \cite{Mokeev:2012vsa}.
Our lattice simulation cannot provide any conclusion regarding this scenario since we have
not incorporated interpolating fields of the hybrid type.

\subsection{Other possibilities for absence of the resonance related level}

Let us discuss other possible reasons for the missing resonance level in our results,
beyond the coupled-channel interpretation offered above.

We could be missing the  eigenstate because we might have missed important coupling
operators. One such candidate might be a genuine pentaquark operator. A local five quark
interpolator (with baryon-meson color structure) has been used by \cite{Kiratidis:2016hda}
who, however, also did not find a Roper signal. The  local pentaquark operator with color
structure $\epsilon_{abc} \bar q_a [qq]_b[qq]_c$ ($[qq]_c=\epsilon_{cde} q_cq_dq_e$)   can
be rewritten as a linear combination of local baryon-meson  operators  $BM=(\epsilon_{abc}
q_a q_b q_c) (\bar q_eq_e)$ by using 
$\epsilon_{abc}\epsilon_{ade}=\delta_{bd}\delta_{ce}-\delta_{be}\delta_{cd}$.   
Furthermore,  the local baryon-meson operators  are linear combinations of $B({\mathbf
p})M(-\mathbf{p})$. Among various terms, the  $N(1)\pi(-1)$  and $N(0)\sigma(0)$ are the
essential ones for the explored energy region and those were incorporated in our basis
(\ref{O}). So, we expect  that our simulation does incorporate the most essential
operators in the linear combination representing  the genuine localized pentaquark
operator.   It remains to be seen if structures with significantly separated  diquark
(such as proposed  in \cite{Lebed:2015tna} for $P_c$) could be also be probed by
baryon-meson operators like (\ref{O}).

It could also be that -- contrary to our expectation -- using operators with
different quark smearing widths is not sufficient to scan the $qqq$ radial excitations.
One might have to expand the interpolator set to include non-local interpolators
\cite{Edwards:2011jj} so as to have good overlap with radial excitations with non-trivial
nodal structures. There has been no study that involved use of such operators along with
the baryon-meson operators and within the single hadron approach such operators do not
produce low lying levels in the Roper energy range  \cite{Edwards:2011jj}.

Finally,  our results are obtained using fermions that do not obey exact chiral symmetry
at finite lattice spacing $a$, like in most of the  previous simulations. It would be 
desirable to verify our results   using fermions that respect  chiral symmetry at finite
$a$.

\section{Conclusion and outlook}
 
We have determined the spectrum of the $J^P=1/2^+$ and $I=1/2$ channel below 1.65 GeV,
where the Roper resonance appears in experiment.  This lattice simulation has been
performed on the PACS-CS ensemble with $N_f=2+1$, $m_\pi\simeq 156$ MeV and $L=2.9~$fm. 
Several interpolating fields of type $qqq$ ($N$) and $qqqq\bar q$ ($N\sigma$ in $s$-wave
and $N\pi$ in $p$-wave) were incorporated, and three eigenstates below $1.65~$GeV are
found. The  energies, their overlaps to the interpolating fields and additional arguments
presented in the paper indicate that these are related to the states that correspond to 
$N(0)$, $N(0)\pi(0)\pi(0)$ and $N(1)\pi(-1)$ in the non-interacting limit (momenta in
units of $2\pi/L$ are given in parenthesis). This is the first simulation that finds the
expected multi-hadron states in this channel. However, the uncertainties on the extracted
energies are sizable and the extracted $N\pi$ phase shift is consistent with zero within a
large error.

One of our main results is that  only three eigenstates lie below $1.65~$GeV, while the
fourth one lies already at about $1.8(1)~$GeV or higher. In contrast, the experimental
$N\pi$ phase shift implies four lattice energy levels below 1.65 GeV    in the elastic
approximation when $N\pi$ is decoupled from $N\pi\pi$ and the later channel is
non-interacting. Our results indicate that the low-lying Roper resonance does not arise on
the lattice within the elastic approximation of $N\pi$ scattering. This  points to a
possibility of a dynamically generated resonance, where  the coupling of $N\pi$ with
$N\pi\pi$ or other channels is essential for the existence of this resonance. This is
supported by comparable overlaps of the operator $O^{N\sigma}$ to the second and third
eigenstates.

We come to a similar conclusion if we compare our lattice spectrum to the HEFT predictions
for  $N\pi /N\sigma /\Delta\pi$ scattering in three scenarios \cite{Liu:2016uzk}. The case
where these three channels are coupled with the low-lying bare Roper $qqq$  core is
disfavored. Our results favor the  scenario where the Roper resonance arises solely  as a
coupled channel phenomenon, without the  Roper $qqq$ core.

Future steps  towards a better understanding of this channel include  simulations at
larger $m_\pi L$, decreasing the statistical error and employing $qqq$ or $qqqq\bar q$
operators with  greater variety of  spatially-extended structures. Simulating the system
at non-zero total momentum will give further information but will introduce additional
challenges: states of positive as well as negative parity contribute to the relevant
irreducible representations in this case.    It would also be important to investigate the
spectrum  based on  fermions with exact chiral symmetry at finite lattice spacing.

Our results point towards the possibility that Roper resonance is a coupled-channel
phenomenon. If this is the case,  the rigorous treatment of this  channel on the lattice
will be  challenging. This is due to  the three-hadron decay channel $N\pi\pi$  and the
fact that the three-hadron scattering matrix has never been extracted from lattice QCD
calculations yet.  The simplified two-body  approach to coupled-channels $N\sigma
/\Delta\pi$ (based on stable $\sigma$ and $\Delta$)  cannot  be compared quantitatively 
to the lattice data at light $m_\pi$ where $\sigma$ and $\Delta$ are broad unstable
resonances.  This is manifested  also in our simulation, where $O^{N\sigma}$ operator
renders an eigenstate with $E\simeq m_N+2m_\pi$ and not $E\simeq m_N+m_\sigma$.

Pion-nucleon scattering has been the prime source of our present day knowledge on hadrons.
After decades of lattice QCD calculations we are now approaching the possibility to study
that scattering process from first principles. This has turned out to be quite challenging
and our contribution is only one step of more to follow.  
 
 \acknowledgements
 
We thank the PACS-CS collaboration for providing the gauge configurations.  We would
kindly like to thank M. D{\"o}ring,   L. Glozman,  Keh-Fei Liu and D. Mohler    for
valuable discussions. We are grateful to B. Golli,  M. Rosina and S. \v Sirca for careful
reading of the manuscript and numerous valuable discussions and suggestions. This work is
supported in part by the  Slovenian Research Agency ARRS, by the Austrian Science Fund
FWF:I1313-N27 and by  the Deutsche Forschungsgemeinschaft Grant No. SFB/TRR 55. The
calculations were performed on  computing clusters at  the University of Graz (NAWI Graz)
and Ljubljana.  S.P. acknowledges support from U.S. Department of Energy Contract No.
DE-AC05-06OR23177, under which Jefferson Science Associates, LLC, manages and operates
Jefferson Lab. 
               

 \appendix
     
\section{An example of a Wick contraction  }\label{sec:wick}
 
Here an example   of a Wick contraction is sketched in order to illustrate how one deals
with the spin  components at the source and sink. Let us consider the correlation function
for the first   $n\pi^+$ term in  $O^{N\pi,\;m_s=1/2}$ at the sink and   $O^{N,\;m_s=1/2}$
at the source (\ref{O})
   \begin{align}
   &\langle n_{-1/2}(-e_x)\pi^+(e_x)|p_{1/2}(0)\rangle =\\ 
  =&\langle (u^T \Gamma_2 d) (\Gamma_1 d)_{\mu=2}~(\bar d\gamma_5 u)\; |\; ( \bar u \Gamma_1^\prime )_{\mu^\prime =1} (\bar d \Gamma_2^\prime \bar u)\rangle\nonumber \\
 =&\langle u_{\alpha} (\Gamma_2)_{\alpha\beta} d_\beta~ (\Gamma_1 d)_{\mu}~\bar d_\gamma (\gamma_5)_{\gamma\delta} u_\delta\; |\; ( \bar u\Gamma_1^\prime)_{\mu^\prime } ~\bar d_{\alpha^\prime}  (\Gamma_2^\prime)_{\alpha^\prime\beta^\prime} \bar u_{\beta^\prime}\rangle\nonumber \\
=& - (\Gamma_1 d\bar d)_{\mu\gamma} (\Gamma_2)_{\alpha\beta}(\Gamma_2^\prime)_{\alpha^\prime\beta^\prime}(\gamma_5)_{\gamma\delta} (d\bar d)_{\beta \alpha^\prime}(u\bar u)_{\delta \beta^\prime}  (u\bar u\Gamma_1^\prime)_{\alpha \mu^\prime} \nonumber\\
&+\mathrm{three\ contr.}\nonumber\\
=&M_{\mu\mu^\prime}+\mathrm{three\ contr.}=M_{21}+\mathrm{three\ contr.}\nonumber
   \end{align}
Among four Wick contractions one is shown as an example: there $\bar d$ from the
pion at the sink contracts with $(\Gamma_1 d)_{\mu}$ from the neutron at the sink, while
the remaining quark lines follow a standard proton contraction. All indices except for
Dirac indices are omitted for simplicity.

The open Dirac-spinor index  is $\mu^\prime=1$ at the source and $\mu=2$ at the sink for
this particular term, while all other Dirac indices are summed over. The open indices
$\mu$ and $\mu^\prime$  can be represented in the matrix form   $M_{\mu\mu^\prime}$ where
the element $M_{21}$ is relevant for the given contraction. Any Wick contraction   in our
correlation matrix can be represented by some matrix $M_{\mu\mu^\prime}$, where 
$\mu^\prime=1$ ($\mu^\prime=2$) is taken  for nucleon with spin up (down) in the source,
and $\mu=1$ ($\mu=2$)   for nucleon with spin up (down) in the sink.

\section{More on the effective energies}\label{app:effmasses}
       
\begin{figure*}[!htb]
\begin{center} 
\includegraphics[scale=0.57]{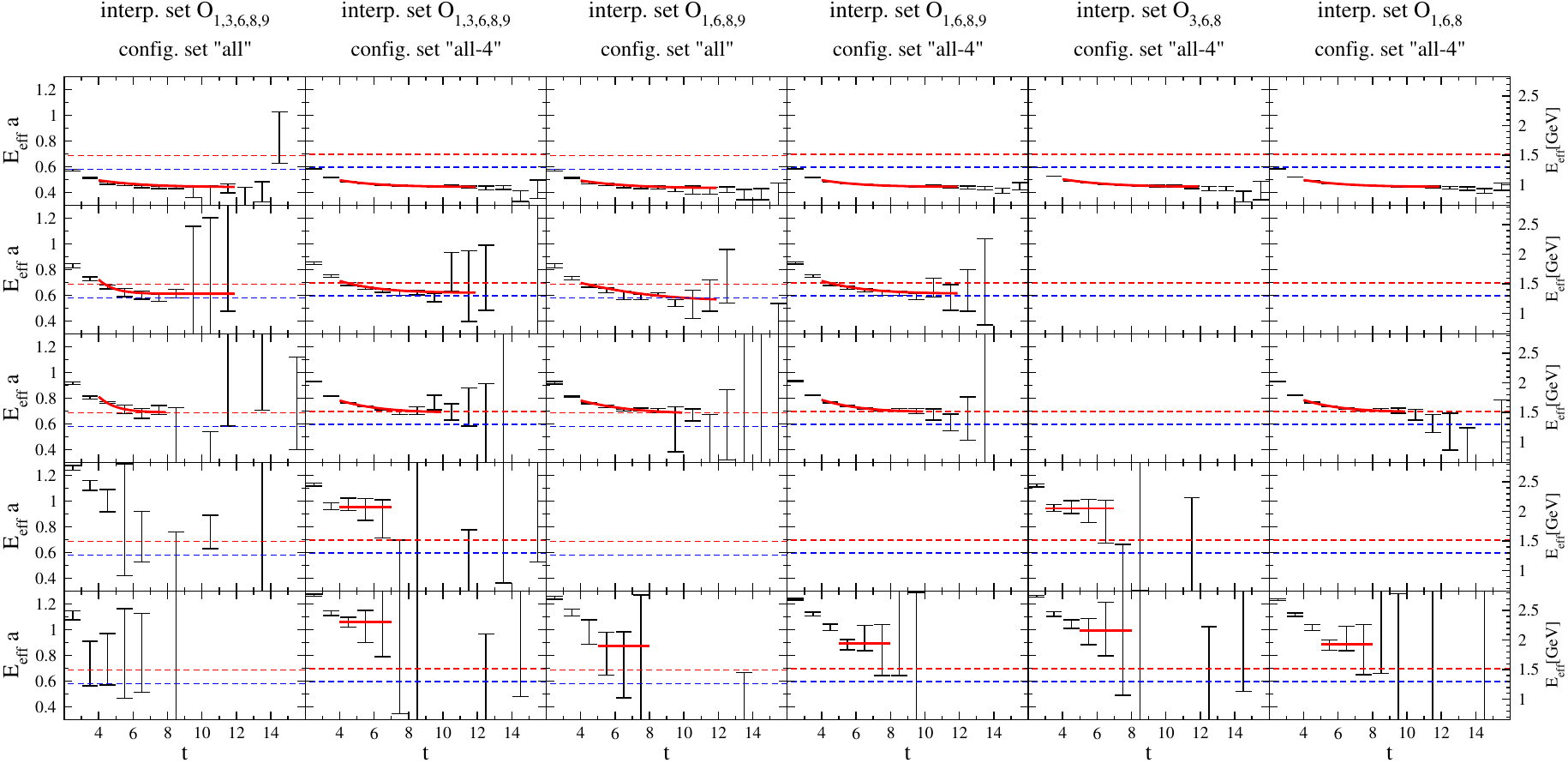}
\end{center}
\caption{Effective energies $E_n^{eff}(t)$ for various choices of interpolator sets and
configuration sets, that are discussed in Section \ref{sec:conf}. The dashed horizontal
lines present non-interacting energies of $N(0)\pi(0)\pi(0)$ (blue dashed) and
$N(1)\pi(-1)$ (red dashed) for the corresponding configuration sets. The fit estimates are
shown as red solid curves. The highest energy levels lie near or above 2 GeV and we refrain from fitting those since no clear plateau is observed.}
\label{fig:Eeff_set}
\end{figure*}

The effective energies for various choices of interpolator and configuration sets are
presented in Fig. \ref{fig:Eeff_set}. 
               
\end{document}